\documentclass[12pt]{article}
\renewcommand{\thefootnote}{\fnsymbol{footnote}}
\newlength{\pubnumber} 

\catcode`\@=11
\@addtoreset{equation}{section}

\def\section{\@startsection{section}{1}{\z@}{3.5ex plus 1ex minus .2ex}
 {2.3ex plus .2ex}{\large\bf}}
\def\subsection{\@startsection{subsection}{2}{\z@}{2.3ex plus .2ex}
 {2.3ex plus .2ex}{\bf}}

\usepackage{amssymb}
\usepackage{epsfig}
\usepackage{cite}
\begin{document}
\begin{titlepage}
\samepage{
\setcounter{page}{1}
\rightline{LTH--714}
\rightline{\tt hep-th/0610118}
\vfill
\begin{center}
 {\Large \bf  Minimal Standard Heterotic String Models \\}
\vfill
\vfill
 {\large  Alon E. Faraggi\footnote{
        E-mail address: faraggi@amtp.liv.ac.uk},
Elisa Manno\footnote{
	E-mail address: Elisa.Manno@liv.ac.uk}
and
	  Cristina Timirgaziu$\setcounter{footnote}{3}$\footnote{
        E-mail address: timirgaz@amtp.liv.ac.uk}\\
}
\vspace{.12in}
 {\it           Department of Mathematical Sciences,
		University of Liverpool,
                Liverpool L69 7ZL}
\end{center}
\vfill
\begin{abstract}
Three generation heterotic--string vacua in the free fermionic formulation
gave rise to models with solely the MSSM states in the observable Standard
Model charged sector. The relation of these models to $Z_2\times Z_2$ 
orbifold compactifications dictates that they produce
three pairs of untwisted Higgs multiplets. The reduction to one
pair relies on the analysis of supersymmetric flat directions,
that give superheavy mass to the dispensable Higgs states. We
explore the removal of the extra Higgs representations by using the 
free fermion boundary conditions and hence directly at the string level, rather
than in the effective low energy field theory. We present a general 
mechanism that achieves this reduction by using asymmetric
boundary conditions between the left-- and right--moving internal fermions. 
We incorporate this mechanism in explicit string models containing 
three twisted generations and a single untwisted Higgs doublet pair.
We further demonstrate that an additional effect of
the asymmetric boundary conditions is to substantially reduce the
supersymmetric moduli space. 

\end{abstract}
\smallskip}
\end{titlepage}

\renewcommand{\thefootnote}{\arabic{footnote}}
\setcounter{footnote}{0}

\def\l{\label}
\def\beq{\begin{equation}}
\def\eeq{\end{equation}}
\def\beqn{\begin{eqnarray}}
\def\eeqn{\end{eqnarray}}

\def\ie{{\it i.e.}}
\def\eg{{\it e.g.}}
\def\half{{\textstyle{1\over 2}}}
\def\third{{\textstyle {1\over3}}}
\def\quarter{{\textstyle {1\over4}}}
\def\m{{\tt -}}
\def\p{{\tt +}}

\def\slash#1{#1\hskip-6pt/\hskip6pt}
\def\slk{\slash{k}}
\def\GeV{\,{\rm GeV}}
\def\TeV{\,{\rm TeV}}
\def\y{\,{\rm y}}
\def\SM{Standard-Model }
\def\SUSY{supersymmetry }
\def\SSSM{supersymmetric standard model}
\def\vev#1{\left\langle #1\right\rangle}
\def\l{\langle}
\def\r{\rangle}

\def\Htw{{\tilde H}}
\def\chibar{{\overline{\chi}}}
\def\qbar{{\overline{q}}}
\def\ibar{{\overline{\imath}}}
\def\jbar{{\overline{\jmath}}}
\def\Hbar{{\overline{H}}}
\def\Qbar{{\overline{Q}}}
\def\abar{{\overline{a}}}
\def\alphabar{{\overline{\alpha}}}
\def\betabar{{\overline{\beta}}}
\def\tautwo{{ \tau_2 }}
\def\thetatwo{{ \vartheta_2 }}
\def\thetathree{{ \vartheta_3 }}
\def\thetafour{{ \vartheta_4 }}
\def\ttwo{{\vartheta_2}}
\def\tthree{{\vartheta_3}}
\def\tfour{{\vartheta_4}}
\def\ti{{\vartheta_i}}
\def\tj{{\vartheta_j}}
\def\tk{{\vartheta_k}}
\def\calF{{\cal F}}
\def\smallmatrix#1#2#3#4{{ {{#1}~{#2}\choose{#3}~{#4}} }}
\def\ab{{\alpha\beta}}
\def\Minv{{ (M^{-1}_\ab)_{ij} }}
\def\bone{{\bf 1}}
\def\ii{{(i)}}
\def\V{{\bf V}}
\def\b{{\bf b}}
\def\N{{\bf N}}
\def\t#1#2{{ \Theta\left\lbrack \matrix{ {#1}\cr {#2}\cr }\right\rbrack }}
\def\C#1#2{{ C\left\lbrack \matrix{ {#1}\cr {#2}\cr }\right\rbrack }}
\def\tp#1#2{{ \Theta'\left\lbrack \matrix{ {#1}\cr {#2}\cr }\right\rbrack }}
\def\tpp#1#2{{ \Theta''\left\lbrack \matrix{ {#1}\cr {#2}\cr }\right\rbrack }}
\def\l{\langle}
\def\r{\rangle}


\def\inbar{\,\vrule height1.5ex width.4pt depth0pt}

\def\IC{\relax\hbox{$\inbar\kern-.3em{\rm C}$}}
\def\IQ{\relax\hbox{$\inbar\kern-.3em{\rm Q}$}}
\def\IR{\relax{\rm I\kern-.18em R}}
 \font\cmss=cmss10 \font\cmsss=cmss10 at 7pt
\def\IZ{\relax\ifmmode\mathchoice
 {\hbox{\cmss Z\kern-.4em Z}}{\hbox{\cmss Z\kern-.4em Z}}
 {\lower.9pt\hbox{\cmsss Z\kern-.4em Z}}
 {\lower1.2pt\hbox{\cmsss Z\kern-.4em Z}}\else{\cmss Z\kern-.4em Z}\fi}

\def\AEF{A.E. Faraggi}
\def\NPB#1#2#3{{Nucl.\ Phys.}\/ {B \bf #1} (#2) #3}
\def\PLB#1#2#3{{Phys.\ Lett.}\/ {B \bf #1} (#2) #3}
\def\PRD#1#2#3{{Phys.\ Rev.}\/ {D \bf #1} (#2) #3}
\def\PRL#1#2#3{{Phys.\ Rev.\ Lett.}\/ {\bf #1} (#2) #3}
\def\PRP#1#2#3{{Phys.\ Rep.}\/ {\bf#1} (#2) #3}
\def\MODA#1#2#3{{Mod.\ Phys.\ Lett.}\/ {\bf A#1} (#2) #3}
\def\IJMP#1#2#3{{Int.\ J.\ Mod.\ Phys.}\/ {A \bf #1} (#2) #3}
\def\nuvc#1#2#3{{Nuovo Cimento}\/ {\bf #1A} (#2) #3}
\def\JHEP#1#2#3{{JHEP} {\textbf #1}, (#2) #3}
\def\etal{{\it et al\/}}

\newcommand{\be}{\begin{equation}}
\newcommand{\ee}{\end{equation}}
\newcommand{\ba}{\begin{eqnarray}}
\newcommand{\ea}{\end{eqnarray}}
\hyphenation{su-per-sym-met-ric non-su-per-sym-met-ric}
\hyphenation{space-time-super-sym-met-ric}
\hyphenation{mod-u-lar mod-u-lar--in-var-i-ant}


\setcounter{footnote}{0}
\section{Introduction}
\bigskip
String theory provides a viable framework to probe the unification of
gravity and the gauge interactions. Preservation of classical symmetries
in the quantised string fixes the number of world--sheet degrees of freedom
required for internal consistency. These can be taken as
bosonic degrees of freedom and interpreted as additional
dimensions, beyond the four space--time, or as an internal two dimensional
conformal field theory on the string world--sheet.
String states are obtained by acting on the nondegenerate
vacuum with bosonic and fermionic excitations and give rise
to the matter and interaction states that constitute the experimentally
observed spectrum.
String theory in four dimensions gives rise to
a multitude of consistent vacuum solutions and the selection of the
ones relevant for experimental investigation is one of the perplexing
issues in string theory.

On the other hand,
the data extracted from collider and other contemporary experiments
highlights the Standard Particle Model as the correct accounting of all the
observed data. Furthermore, the particle physics
data is compatible with the hypothesis that the renormalisable Standard Model
remains unaltered up to a large energy scale and that the particle spectrum
is embedded in a Grand Unified Theory (GUT) \cite{gutsreviews}.
Most appealing in this context is SO(10)
unification, in which each Standard Model generation is embedded in a
single 16 spinorial representation. The hypothesis of unification
is further supported by the observed logarithmic evolution of the
Standard Model parameters, by the suggestive compatibility
of coupling unification with the low energy data and by the suppression
of proton decay mediating operators and left--handed neutrino masses.
 However, as many of the Standard Model variables are mere parameters
in the context of GUTs, their origin is not explained in the framework
of point quantum field theories.
Therefore, elucidating further the properties
of the Standard Model spectrum, such as the existence of flavour, necessitates
the unification of the Standard Particle Model with gravity.
String theory therefore provides a unique framework to explore how the
Standard Model parameters may arise from such unification.

Toward this end and lacking a mechanism that dynamically selects
a unique string vacuum in four dimensions, the Standard Model
data is used to single out phenomenologically viable string
vacua.
Maintaining the SO(10) embedding of the Standard Model spectrum
necessitates that we study compactifications of the heterotic
string \cite{hete}, as spinorial $SO(10)$ representations are obtained
in the heterotic string \cite{hete}, but not in the other
perturbative string theories. The compatibility of the
low energy data with gauge coupling unification indicates
that the observable gauge group below the string scale
should be $SU(3)_C\times SU(2)_L\times U(1)_Y$ and that the
$SU(3)_C\times SU(2)_L\times U(1)_Y$--charged spectrum
should consist solely of the Minimal Supersymmetric Standard
Model spectrum. Indeed, over the past year there has been a resurgence
of interest in the construction of phenomenologically viable
heterotic string vacua \cite{resurgence}. We therefore recall
that the semi--realistic models in the free fermionic formulation
\cite{fsu5,fny,fc,alr,eu,top,nahe,cfn}
produced in the past solutions in which the only Standard Model
charged states are the MSSM states \cite{cfn}. It is therefore an
opportune moment
to revisit these models and to examine some of their properties.

One of the intriguing successes of the free fermionic standard--like models
has been the successful prediction of the top quark mass several years
prior to its experimental observation \cite{top}.
Furthermore, the models offered
an explanation why only the top quark mass is characterised by the
electroweak scale, whereas the masses of the lighter quarks and
leptons are suppressed. The reason being that only the top quark
Yukawa coupling is obtained at the cubic level of the superpotential,
whereas the Yukawa couplings of the
lighter quarks and leptons are obtained from nonrenormalizable terms
which are suppressed relative to the leading order term. In the free
fermionic quasi--realistic standard--like models the three generations
arise from the three twisted sectors, whereas the Higgs doublets, to which
they couple in leading order, arise from the untwisted sector. At leading
order each twisted generation couples to a separate pair of untwisted
Higgs doublets. Analysis of supersymmetric flat directions implied
that at low energies only one pair of Higgs doublets remains light and
other Higgs doublets obtain heavy mass from VEVs of Standard Model singlet
fields. Hence, in the low energy effective field theory, only the
coupling of the twisted generation that couples to the light Higgs
remains at leading order. The consequence is that only the top quark
mass is obtained at leading order, whereas the masses of the remaining
quarks and leptons are obtained at subleading orders. Evolution of the
calculated Yukawa couplings from the string to electroweak scale
(and using the low energy data for the bottom quark and electroweak
vector bosons) then yields a prediction for the top quark mass. An additional
constraint on the analysis is that the gauge coupling unification at the
high scale is compatible with the low energy data.

The analysis of the top quark mass therefore relies on the analysis
of supersymmetric flat directions and the decoupling of the additional
untwisted electroweak Higgs doublets, that couple to the twisted generations
at leading order. In this paper we examine whether an alternative
construction is possible. Our aim is to construct models in which
only one pair of untwisted Higgs doublets remains in the massless
spectrum after application of the Generalised GSO (GGSO) projections.
Therefore, the massless string spectrum contains a single
electroweak Higgs doublet pair, without relying on analysis of supersymmetric
flat direction in the effective low energy field theory.
Consequently, at leading order only the top quark couples to
the electroweak Higgs doublet, and therefore only its mass
is intrinsically associated with the electroweak scale.

Our paper is organised as follows. In section \ref{review} we review the
structure of the minimal standard heterotic string models in the
free fermionic formulation. In section \ref{hdts} we describe
in general the mechanism that projects the additional untwisted Higgs doublets,
in terms of the free fermion boundary condition basis vectors.
In section \ref{higgsreduced}
we present an explicit string model which contains a single untwisted
Higgs doublet pair and present its full massless spectrum as well
as the full cubic level superpotential. Section  \ref{higgsreduced} also contains a discussion of the flat directions. Section \ref{conclude} concludes the
paper.

\section{Minimal Standard Heterotic String Models}\label{review}
In this section we briefly review the construction and structure of the
free fermionic standard like models.
The notation and further details of the construction of these
models are given elsewhere \cite{fny,eu,nahe,cfn,cfs}.
In the free fermionic formulation of the heterotic string
in four dimensions \cite{fff} all the world--sheet
degrees of freedom,  required to cancel
the conformal anomaly, are represented in terms of free fermions
propagating on the string world--sheet.
In the light--cone gauge the world--sheet field content consists
of two transverse left-- and right--moving space--time coordinate bosons,
$X_{1,2}^\mu$ and ${\bar X}_{1,2}^\mu$,
and their left--moving fermionic superpartners $\psi^\mu_{1,2}$,
and additional 62 purely internal
Majorana--Weyl fermions, of which 18 are left--moving,
and 44 are right--moving.
The models are constructed by specifying the phases picked by
the world--sheet fermions when transported along the torus
non--contractible loops. Each model corresponds to a particular choice
of fermion phases consistent with modular invariance and is generated
by a set of basis vectors describing the transformation properties
of the 64 world--sheet fermions.
The physical spectrum is obtained by applying the generalised GSO projections.
The low energy effective field theory is obtained by S--matrix elements
between external states \cite{kln}.

The boundary condition basis defining a typical
realistic free fermionic heterotic string models is
constructed in two stages.
The first stage consists of the NAHE set,
which is a set of five boundary condition basis vectors,
$\{{\bf1},S,b_1,b_2,b_3\}$ \cite{costas,nahe}.
The gauge group after imposing the GSO projections induced
by the NAHE set is $SO(10)\times SO(6)^3\times E_8$
with $N=1$ supersymmetry.
The NAHE set divides the internal world--sheet
fermions in the following way: ${\bar\phi}^{1,\cdots,8}$ generate the
hidden $E_8$ gauge group, ${\bar\psi}^{1,\cdots,5}$ generate the $SO(10)$
gauge group, and $\{{\bar y}^{3,\cdots,6},{\bar\eta}^1\}$,
$\{{\bar y}^1,{\bar y}^2,{\bar\omega}^5,{\bar\omega}^6,{\bar\eta}^2\}$,
$\{{\bar\omega}^{1,\cdots,4},{\bar\eta}^3\}$ generate the three horizontal
$SO(6)$ symmetries. The left--moving $\{y,\omega\}$ states are divided
to $\{{y}^{3,\cdots,6}\}$,
$\{{y}^1,{y}^2,{\omega}^5,{\omega}^6\}$,
$\{{\omega}^{1,\cdots,4}\}$, while $\chi^{12}$, $\chi^{34}$, $\chi^{56}$
generate the left--moving $N=2$ world--sheet supersymmetry.
At the level of the NAHE set the sectors $b_1$, $b_2$ and $b_3$
produce 48 multiplets, 16 from each, in the $16$
representation of $SO(10)$, that are
singlets of the hidden $E_8$ gauge group and transform
under the horizontal $SO(6)_j$ $(j=1,2,3)$ symmetries.
The untwisted sector produces states in the 10 vectorial representation
of $SO(10)$ that can produce electroweak Higgs doublets.
At this stage we anticipate that the $SO(10)$ group produces the
Standard Model gauge group factors, and that the $16\cdot16\cdot10$
can produce the Standard Model fermion mass terms.
This structure
is common to all the realistic free fermionic models that we consider here.

The second stage of the
basis construction consists of adding to the
NAHE set three additional boundary condition basis vectors.
These additional basis vectors reduce the number of generations
to three chiral generations, one from each of the sectors $b_1$,
$b_2$ and $b_3$, and simultaneously break the four dimensional
gauge group. The assignment of boundary conditions to
$\{{\bar\psi}^{1,\cdots,5}\}$ breaks $SO(10)$ to one of its subgroups
$SU(5)\times U(1)$ \cite{fsu5}, $SO(6)\times SO(4)$ \cite{alr},
$SU(3)\times SU(2)\times U(1)^2$ \cite{fny,eu,cfn},
$SU(3)\times SU(2)^2\times U(1)$ \cite{cfs} or
$SU(4)\times SU(2)\times U(1)$ \cite{cfnooij}.
Similarly, the hidden $E_8$ symmetry is broken to one of its
subgroups. The flavour $SO(6)^3$ symmetries in the NAHE--based models
are always broken to flavour $U(1)$ symmetries, as the breaking
of these symmetries is correlated with the number of chiral
generations. Three such $U(1)_j$ symmetries are always obtained
in the NAHE based free fermionic models, from the subgroup
of the observable $E_8$, which is orthogonal to $SO(10)$.
These are produced by the world--sheet currents ${\bar\eta}{\bar\eta}^*$
($j=1,2,3$), which are part of the Cartan sub--algebra of the
observable $E_8$. Additional unbroken $U(1)$ symmetries, denoted
typically by $U(1)_j$ ($j=4,5,...$), arise by pairing two real
fermions from the sets
$\{{\bar y}^{3,\cdots,6}\}$,
$\{{\bar y}^{1,2},{\bar\omega}^{5,6}\}$ and
$\{{\bar\omega}^{1,\cdots,4}\}$.
The final observable gauge
group depends on the number of such pairings.
Alternatively, a left--moving real fermion from the sets
$\{{ y}^{3,\cdots,6}\}$, $\{{ y}^{1,2},{\omega}^{5,6}\}$ and
$\{{\omega}^{1,\cdots,4}\}$ may be paired with its respective
right--moving real fermion to form an Ising model operator,
in which case the rank of the right--moving gauge group is reduced
by one.
The reduction of untwisted
electroweak Higgs doublets crucially depends on the pairings
of the left-- and right--moving fermions from the set
$\{y,\omega|{\bar y},{\bar\omega}\}^{1\cdots6}$.

Subsequent to constructing the basis vectors and extracting the massless
spectrum, the analysis of the free fermionic models proceeds by
calculating the superpotential. The cubic and higher-order terms in
the superpotential are obtained by evaluating the correlators
\beq
A_N\sim \langle V_1^fV_2^fV_3^b\cdots V_N^b\rangle,
\label{supterms}
\eeq
where $V_i^f$ $(V_i^b)$ are the fermionic (scalar) components
of the vertex operators, using the rules given in~\cite{kln}.
Generically, correlators of the form (\ref{supterms}) are of order
${\cal O} (g^{N-2})$, and hence of progressively higher orders
in the weak-coupling limit. Typically,
one of the $U(1)$ factors in the free-fermion models is anomalous
and generates a Fayet--Iliopoulos term which breaks supersymmetry
at the Planck scale \cite{dsw}. A supersymmetric vacuum is obtained by
assigning non--trivial VEVs to a set of Standard Model singlet
fields in the massless string spectrum along $F$ and $D$--flat directions.
Some of these fields will appear in the nonrenormalizable terms
(\ref{supterms}), leading to
effective operators of lower dimension. Their coefficients contain
factors of order ${\cal V} / M{\sim 1/10}$.

An example of free fermionic standard--like model is given in table
\ref{fnymodel}.

\beqn
 &\begin{tabular}{c|c|ccc|c|ccc|c}
 ~ & $\psi^\mu$ & $\chi^{12}$ & $\chi^{34}$ & $\chi^{56}$ &
        $\bar{\psi}^{1,...,5} $ &
        $\bar{\eta}^1 $&
        $\bar{\eta}^2 $&
        $\bar{\eta}^3 $&
        $\bar{\phi}^{1,...,8} $ \\
\hline
\hline
  ${b_4}$     &  1 & 1&0&0 & 1~1~1~1~1 & 1 & 0 & 0 & 0~0~0~0~0~0~0~0 \\
  ${\alpha}$   &  1 & 0&0&1 & 1~1~1~0~0 & 1 & 1 & 0 & 1~1~1~1~0~0~0~0 \\
  ${\beta}$  &  1 & 0&1&0 &
                ${1\over2}$~${1\over2}$~${1\over2}$~${1\over2}$~${1\over2}$
              & ${1\over2}$ & ${1\over2}$ & ${1\over2}$ &
                ${1\over2}$~0~1~1~${1\over2}$~${1\over2}$~${1\over2}$~0 \\
\end{tabular}
   \nonumber\\
   ~  &  ~ \nonumber\\
   ~  &  ~ \nonumber\\
     &\begin{tabular}{c|c|c|c}
 ~&   $y^3{y}^6$
      $y^4{\bar y}^4$
      $y^5{\bar y}^5$
      ${\bar y}^3{\bar y}^6$
  &   $y^1{\omega}^6$
      $y^2{\bar y}^2$
      $\omega^5{\bar\omega}^5$
      ${\bar y}^1{\bar\omega}^6$
  &   $\omega^1{\omega}^3$
      $\omega^2{\bar\omega}^2$
      $\omega^4{\bar\omega}^4$
      ${\bar\omega}^1{\bar\omega}^3$ \\
\hline
\hline
$b_4$ & 1 ~~~ 0 ~~~ 0 ~~~ 1  & 0 ~~~ 0 ~~~ 1 ~~~ 0  & 0 ~~~ 0 ~~~ 1 ~~~ 0 \\
$\alpha$  & 0 ~~~ 0 ~~~ 0 ~~~ 1  & 0 ~~~ 1 ~~~ 0 ~~~ 1  & 1 ~~~ 0 ~~~ 1 ~~~ 0
\\
$\beta$ & 0 ~~~ 0 ~~~ 1 ~~~ 1  & 1 ~~~ 0 ~~~ 0 ~~~ 1  & 0 ~~~ 1 ~~~ 0 ~~~ 0 \\
\end{tabular}
\label{fnymodel}
\eeqn
The choice of generalised GSO coefficients is:
\beqn
&&c\left(\matrix{b_4\cr
                                    b_j,\alpha\cr}\right)=
-c\left(\matrix{b_4\cr
                                    {\bf1}\cr}\right)=
-c\left(\matrix{\alpha\cr
                                    {\bf1}\cr}\right)=
c\left(\matrix{\alpha\cr
                                    b_j\cr}\right)=\nonumber\\
&&-c\left(\matrix{\alpha\cr
                                    \beta\cr}\right)=
c\left(\matrix{\beta\cr
                                    b_2\cr}\right)=
-c\left(\matrix{\beta\cr
                                    b_1,b_3,b_4,\beta\cr}\right)=
-1\nonumber
\eeqn
$(j=1,2,3),$
with the others specified by modular invariance and space--time
supersymmetry. The full massless spectrum, charged under the four
dimensional gauge group of this model was presented
in ref. \cite{fny}, as well as the full cubic level superpotential.
In ref. \cite{fc} it was noted that all the exotic fractionally charged
states in the model decouple from the effective low energy field theory
at the cubic level of the superpotential, provided that the
set of Standard Model singlet fields
${\bar\phi}_4,{\bar\phi}_4',{\phi}_4,\phi_4'$ obtain a string scale VEV.
Supersymmetric flat solutions that incorporate these VEVs were found
in ref. \cite{cfn}, given by the VEVs of the set of fields
\beq
\{
\phi_{12}, \phi_{23}, {\bar\phi}_{56}, \phi_4, \phi_4^\prime,
{\bar\phi}_4,{\bar\phi}_4^\prime, H_{15}, H_{30},
H_{31}, H_{38} \}
\label{fdsol}.
\eeq
Additionally it was demonstrated in \cite{cfn} that in this vacuum
solution the only Standard Model charged states that remain massless below
the anomalous $U(1)$ scale consist of the states of the minimal supersymmetric
standard model. Such solutions are therefore dubbed Minimal Standard
Heterotic String Models (MSHSM).

\subsection{Yukawa Selection Mechanism}\label{ysm}

At the cubic level of the superpotential
the boundary condition
basis vectors fix the cubic level Yukawa couplings for the quarks
and leptons \cite{top}.
These Yukawa couplings are fixed by the vector $\gamma$
which breaks the $SO(10)$ symmetry to $SU(5)\times U(1)$.
Each sector
$b_i$ gives rise to an up--like or down--like cubic level Yukawa coupling.
We can define a quantity $\Delta$ in the vector $\gamma$,
which measures the difference between
the left-- and right--moving
boundary conditions assigned to the internal fermions from the set
$\{y,w\vert{\bar y},{\bar\omega}\}$ and which are periodic in the vector
$b_i$,
\begin{equation}
\Delta_i=\vert\gamma_L({\rm internal})-
\gamma_R({\rm internal})\vert=0,1~~(i=1,2,3)
\label{udyc}.
\end{equation}
If $\Delta_i=0$ then the sector $b_i$ gives rise to a
down--like Yukawa coupling while the
up--type Yukawa coupling vanishes. The opposite occurs if $\Delta_i=1$.
In the model of table [\ref{fnymodel}] the basis vector responsible for the
breaking of $SO(10)$ symmetry to $SU(5)\times U(1)$ is $\beta$ and, therefore,
we obtain at the cubic level
$Q_1u_1{\bar h}_1$, $L_1N_1{\bar h}_1$,
$Q_2d_2{h}_2$, $L_2e_2{h}_2$ and
$Q_3d_3{h}_3$, $L_3e_3{h}_3$,
irrespective of the choice of  GSO projection coefficients.
In models that produce $\Delta_i=1$ for $i=1,2,3$
the down--quark type cubic--level Yukawa couplings vanish
and the models produce only up--quark type Yukawa couplings
at the cubic level of the superpotential. Models with these characteristics
were presented in refs. \cite{eu,top}.

\subsection{Higgs Doublet--Triplet Splitting}\label{hdts}

The Higgs doublet--triplet splitting operates as follows \cite{ps}.
The Neveu--Schwarz sector gives rise to three fields in the
10 representation of $SO(10)$.  These contain the  Higgs electroweak
doublets and colour triplets. Each of those is charged with respect to one
of the horizontal $U(1)$ symmetries $U(1)_{1,2,3}$.  Each one of these
multiplets is associated, by the horizontal symmetries, with one of the
twisted sectors, $b_1$, $b_2$ and $b_3$. The doublet--triplet
splitting results from the boundary condition basis vectors which break
the $SO(10)$ symmetry to $SO(6)\times SO(4)$. We can define a quantity
$\Delta_i$ in these basis vectors which measures the difference between the
boundary conditions assigned to the internal fermions from the set
$\{y,w\vert{\bar y},{\bar\omega}\}$ and which are periodic in the vector
$b_i$,
\begin{equation}
\Delta_i=\vert\alpha_L({\rm internal})-
\alpha_R({\rm internal})\vert=0,1~~(i=1,2,3)
\label{dts}.
\end{equation}
If $\Delta_i=0$ then the Higgs triplets, $D_i$ and ${\bar D}_i$,
remain in the massless spectrum while the Higgs doublets, $h_i$ and ${\bar
h}_i$ are projected out
and the opposite occurs for $\Delta_i=1$.

The rule in Eq. (\ref{dts}) is a generic rule that operates in NAHE--based
free fermionic models.
The model
of table [\ref{fnymodel}] illustrates this rule.
In this model the basis vector that breaks $SO(10)$ symmetry to $SO(6)\times
SO(4)$ is $\alpha$ and, with respect to $\alpha$,
$\Delta_1=\Delta_2=\Delta_3=1$. Therefore,
this model produces three pairs of electroweak
Higgs doublets from the Neveu--Schwarz sector, $h_1$, $\bar h_1$
$h_2$, $\bar h_2$ and $h_3$, $\bar h_3$, and all the untwisted colour
triplets are projected out. Note also that the vector basis $b_4$ is symmetric
with respect to the internal fermions
that are periodic in the vectors $b_i, \ i=1,2,3$ and, therefore, does not
project out the fields in the 10 representation of $SO(10)$.

Another relevant question with regard to the Higgs doublet--triplet
splitting mechanism is whether it is possible to construct models in which
both the Higgs colour triplets and electroweak doublets from the
Neveu--Schwarz sector are projected out by the GSO projections.
This is a viable possibility as we can choose for example
$$\Delta_j^{(\alpha)}=1 ~{\rm and}~ \Delta_j^{(\beta)}=0,$$
where $\Delta^{(\alpha,\beta)}$ are the projections due
to the basis vectors $\alpha$ and $\beta$ respectively.
This is a relevant question as the number of Higgs representations,
which generically appear in the massless spectrum,
is larger than what is allowed by the low energy phenomenology.
Attempts to construct such models were discussed in ref. \cite{ffl2}.
However, in the models presented there this was achieved at the
expense of projecting some of the states from the sectors $b_1$, $b_2$ and
$b_3$. In section \ref{higgsreduced} we present for the first time
three generation models with reduced untwisted Higgs spectrum,
without resorting to analysis of supersymmetric flat directions.

\section{Models with reduced untwisted Higgs spectrum}\label{higgsreduced}
As an illustration of the Higgs reduction mechanism we consider the
model in table \ref{firstexample}
\beqn
 &\begin{tabular}{c|c|ccc|c|ccc|c}
 ~ & $\psi^\mu$ & $\chi^{12}$ & $\chi^{34}$ & $\chi^{56}$ &
        $\bar{\psi}^{1,...,5} $ &
        $\bar{\eta}^1 $&
        $\bar{\eta}^2 $&
        $\bar{\eta}^3 $&
        $\bar{\phi}^{1,...,8} $ \\
\hline
\hline
  ${\alpha}$  &  1 & 1&0&0 & 1~1~1~0~0 & 0 & 1 & 0 & 0~1~1~0~0~0~0~0 \\
  ${\beta}$   &  1 & 0&1&0 & 1~1~1~0~0 & 1 & 1 & 1 & 0~1~1~0~0~0~0~0 \\
  ${\gamma}$  &  1 & 0&0&1 &
		${1\over2}$~${1\over2}$~${1\over2}$~${1\over2}$~${1\over2}$
	      & ${1\over2}$ & ${1\over2}$ & ${1\over2}$ &
                ${1\over2}$~0~0~0~${1\over2}$~${1\over2}$~${1\over2}$~0 \\
\end{tabular}
   \nonumber\\
   ~  &  ~ \nonumber\\
   ~  &  ~ \nonumber\\
     &\begin{tabular}{c|c|c|c}
 ~&   $y^3{y}^6$
      $y^4{\bar y}^4$
      $y^5{\bar y}^5$
      ${\bar y}^3{\bar y}^6$
  &   $y^1{\omega}^6$
      $y^2{\bar y}^2$
      $\omega^5{\bar\omega}^5$
      ${\bar y}^1{\bar\omega}^6$
  &   $\omega^1{\omega}^3$
      $\omega^2{\bar\omega}^2$
      $\omega^4{\bar\omega}^4$
      ${\bar\omega}^1{\bar\omega}^3$ \\
\hline
\hline
$\alpha$ & 1 ~~~ 0 ~~~ 0 ~~~ 0  & 0 ~~~ 0 ~~~ 1 ~~~ 1  & 0 ~~~ 0 ~~~ 1 ~~~ 0 \\
$\beta$  & 0 ~~~ 0 ~~~ 1 ~~~ 1  & 1 ~~~ 0 ~~~ 0 ~~~ 1  & 0 ~~~ 1 ~~~ 0 ~~~ 1 \\
$\gamma$ & 0 ~~~ 1 ~~~ 0 ~~~ 1  & 0 ~~~ 1 ~~~ 0 ~~~ 0  & 1 ~~~ 0 ~~~ 0 ~~~ 1 \\
\end{tabular}
\label{firstexample}
\eeqn
With the choice of generalised GSO coefficients:
\beqn
&& c\left(\matrix{\alpha,\beta\cr
                           \alpha\cr}\right)=
c\left(\matrix{\beta,\gamma\cr
                           \beta\cr}\right)=
-c\left(\matrix{\gamma\cr
                           {\bf1},\alpha\cr}\right)=
c\left(\matrix{\alpha\cr
                                    b_3\cr}\right)=\nonumber\\
&&c\left(\matrix{\gamma\cr
                                    b_1\cr}\right)=
-c\left(\matrix{\beta\cr
                                    b_j\cr}\right)=
-c\left(\matrix{\alpha\cr
                                    b_1,b_2\cr}\right)=
-c\left(\matrix{\gamma\cr
                                   b_2,b_3\cr}\right)=1\nonumber
\eeqn
(j=1,2,3), with the others specified by modular invariance and space--time
supersymmetry. As noted from the table, in this model the boundary conditions
with respect to $b_2$ and $b_3$ in the basis vector $\alpha$
are asymmetric and symmetric, respectively, while the opposite occurs for the basis vector $\beta$.
At the same time, the boundary conditions with respect to the sector
$b_1$ are asymmetric in both $\alpha$ and $\beta$. Therefore,
in this model
$\Delta_1^{(\alpha)}=\Delta_1^{(\beta)}=1$;
$~\Delta_2^{(\alpha)}=1,~\Delta_2^{(\beta)}=0$ and
$\Delta_3^{(\alpha)}=0,~\Delta_3^{(\beta)}=1$.
Consequently, in this model, irrespective of the choice of the generalised GSO
projection coefficients, both the Higgs colour triplets and electroweak
doublets associated with $b_2$ and $b_3$ are projected out by the
GSO projections, whereas the electroweak Higgs doublets that are associated
with the sector $b_1$ remain in the spectrum.
However, in this model, the sector $\alpha$ produces chiral fractionally
charged
exotics, and is therefore not viable. We also note that in this model the
non--vanishing cubic level Yukawa couplings produce a down--quark type mass
term, and not a potential top--quark mass term.

An alternative model is presented in table \ref{secondexample}.
\beqn
 &\begin{tabular}{c|c|ccc|c|ccc|c}
 ~ & $\psi^\mu$ & $\chi^{12}$ & $\chi^{34}$ & $\chi^{56}$ &
        $\bar{\psi}^{1,...,5} $ &
        $\bar{\eta}^1 $&
        $\bar{\eta}^2 $&
        $\bar{\eta}^3 $&
        $\bar{\phi}^{1,...,8} $ \\
\hline
\hline
  $b_4$  &  1 & 1&0&0 & 1~1~1~1~1 & 0 & 1 & 0 & 1~1~1~1~0~0~0~0 \\
  ${\beta}$   &  1 & 0&1&0 & 1~1~1~0~0 & 1 & 1 & 1 & 0~0~0~0~1~1~0~0 \\
  ${\gamma}$  &  1 & 0&0&1 &
		${1\over2}$~${1\over2}$~${1\over2}$~${1\over2}$~${1\over2}$
	      & ${1\over2}$ & ${1\over2}$ & ${1\over2}$ &
                0~0~${1\over2}$~${1\over2}$~0~0~${1\over2}$~${1\over2}$ \\
\end{tabular}
   \nonumber\\
   ~  &  ~ \nonumber\\
   ~  &  ~ \nonumber\\
     &\begin{tabular}{c|c|c|c}
 ~&   $y^3{y}^6$
      $y^4{\bar y}^4$
      $y^5{\bar y}^5$
      ${\bar y}^3{\bar y}^6$
  &   $y^1{\omega}^6$
      $y^2{\bar y}^2$
      $\omega^5{\bar\omega}^5$
      ${\bar y}^1{\bar\omega}^6$
  &   $\omega^1{\omega}^3$
      $\omega^2{\bar\omega}^2$
      $\omega^4{\bar\omega}^4$
      ${\bar\omega}^1{\bar\omega}^3$ \\
\hline
\hline
$b_4$ & 1 ~~~ 0 ~~~ 0 ~~~ 0  & 0 ~~~ 0 ~~~ 1 ~~~ 1  & 0 ~~~ 0 ~~~ 1 ~~~ 0 \\
$\beta$  & 0 ~~~ 0 ~~~ 1 ~~~ 1  & 1 ~~~ 0 ~~~ 0 ~~~ 1  & 0 ~~~ 1 ~~~ 0 ~~~ 1 \\
$\gamma$ & 0 ~~~ 1 ~~~ 0 ~~~ 1  & 0 ~~~ 0 ~~~ 0 ~~~ 1  & 1 ~~~ 1 ~~~ 0 ~~~ 0 \\
\end{tabular}
\label{secondexample}
\eeqn
With the choice of generalised GSO coefficients:
\beqn
&& c\left(\matrix{b_4\cr
                           b_4,\beta,\gamma\cr}\right)=
c\left(\matrix{\beta\cr
                           \beta,\gamma\cr}\right)=
c\left(\matrix{b_4,\gamma\cr
                                    b_j\cr}\right)=\nonumber\\
&& -c\left(\matrix{\gamma\cr
                                   {\bf1}\cr}\right)=
 -c\left(\matrix{\beta\cr
                                   {b_j}\cr}\right)=1\nonumber
\eeqn
(j=1,2,3), with the others specified by modular invariance and space--time
supersymmetry. In this model the basis vector $b_4$ preserves the $SO(10)$
symmetry, which is broken by the basis vectors $\beta$ and $\gamma$ to
$SU(3)\times SU(2)\times U(1)^2$. The $b_4$ projection is asymmetric
with respect to the internal fermions that are periodic in
the sectors $b_1$ and $b_2$ and, therefore, projects out the entire
untwisted vectorial representations of $SO(10)$,
that couple to the sectors $b_1$ and $b_2$, irrespective of the
$\beta$ projection. On the other hand, it is symmetric with respect to $b_3$,
while the basis
vector $\beta$, that breaks $SO(10)\rightarrow SO(6)\times SO(4)$,
is asymmetric with respect to $b_3$. Therefore, the Higgs doublets that
couple to $b_3$ remain in the massless spectrum.
We note also that the boundary conditions
in the vector $\gamma$, that breaks $SO(10)\rightarrow SU(5)\times U(1)$, are
asymmetric with respect to the internal fermions that are periodic in the
sector $b_3$. Therefore, this model will select an up--quark type Yukawa
couplings at the cubic level of the superpotential.
The gauge group of this model is generated entirely from the untwisted vector
bosons and there is no gauge symmetry enhancement from additional sectors.
The four dimensional gauge group is
$SU(3)_C\times SU(2)_L\times U(1)_{B-L}\times U(1)_{T_{3_R}}\times
U(1)_{1,\cdots,6}\times SU(2)_{1,\cdots,6}\times U(1)_{7,8}$.

The spectrum of the model is detailed in the table \ref{matter1} at the end of this paper.
The cubic level superpotential, including states from the observable and hidden
sectors, is straightforwardly calculated following the rules given in
\cite{kln} and reads:
\ba
W&=&N^c_{L_3} L_3\bar{h}+u^c_{L_3} Q_3\bar{h} +
C_+^{-+}D_-\bar{h}+C_-^{+-}D_+h+\nonumber\\
 &+&(\phi_1{\phi_3}'+\phi_1'\phi_3)\phi_2+
(C_+^{-+}C_-^{+-}+C_-^{-+}C_+^{+-})\phi_{3}'\nonumber\\
&+& (D_+D_-+C_+C_-+T_+T_-+D_{+-}^{(6)}D_{-+}^{(6)}+
D_{--}^{(6)}D_{++}^{(6)}){\phi_{3}}\nonumber\\
&+&(D^{(3,4)}_{+-}D^{(3,4)}_{-+}+D_+^{(5)}D_-^{(5)}+
D_{++}^{(3)}D_{--}^{(3)}+D_{+-}^{(3)}D_{-+}^{(3)}){\phi_{1}}
\nonumber\\
&+&A_+A_-\phi_{1}'.\nonumber
\ea

As expected, we obtain a Yukawa coupling for the top quark,
but also couplings of the Higgs with exotic states.
 One can also see that not all the fractionally charged$\footnotemark[1]$
$\footnotetext[1]{The hypercharge is defined as $Q_Y=1/3~Q_C+1/2~Q_L$ and
the electric charge is given by $Q_e=T_{3L}+Q_Y$, with $T_{3L}$ the
electroweak isospin.}$states in the spectrum appear in the cubic level
superpotential, which means that they remain massless at the trilinear level.
However,
this does not exclude the possibility of giving them masses at higher orders.

\subsection{Flat directions}

In this section we investigate the flat directions of the model 
of table \ref{secondexample}. The model contains 6 anomalous $U(1)$'s with
\ba &&{\rm Tr} \ Q_1={\rm Tr}\ Q_2=-{\rm Tr}\ Q_3={\rm Tr}\ Q_5=-24,\nonumber\\
&&{\rm Tr}\ Q_4=-{\rm Tr}\ Q_6=12.\ea
The total anomaly can be rotated into a single $U(1)_A$ and the new basis reads
\ba Q_1'&=&Q_1-Q_2,\nonumber\\
Q_2'&=&Q_3+Q_5,\nonumber\\
Q_3'&=&Q_4+Q_6,\nonumber\\
Q_4'&=&Q_1+Q_2+Q_3-Q_5,\nonumber\\
Q_5'&=&Q_1+Q_2-Q_3+Q_5+4(Q_4-Q_6),\nonumber\\
Q_A&=&2(Q_1+Q_2-Q_3+Q_5)-Q_4+Q_6.
\ea
In the following we will call $Q_i'$, i=1,...,5,  simply $Q_i$.

To search for flat directions we use the methodology developed in
\cite{Cleaver:1998sm}.
We start  by constructing a basis of D-flat directions
under $Q_{1...5}$ and then we investigate the existence of D-flat directions
in the anomalous $U(1)_A$. Subsequently we will have to impose D-flatness
under the remaining gauge groups and F-flatness. To generate the basis of
flat directions under $Q_{1...5}$ we start by forming a basis of gauge
invariant monomials under $U(1)_1$, then we use these invariants to
construct a basis of invariant monomials under $U(1)_2$ and so forth.

We include in the analysis only the fields with vanishing
hypercharge$\footnotemark[1]$ and which are singlets under the
Standard Model gauge group.  The $Q_{1...5,A}$ charges of these fields
are detailed in table \ref{flat},
where, following the notation of \cite{Cleaver:1998sm}, we signal by
$^{(')}$($^{('')}$) the presence in the spectrum of a second (third)
field with the same $U(1)_{1...5,A}$ charges and by $\surd$ the presence
of a field with opposite $U(1)_{1...5,A}$ charges.
For instance, the field $\phi$ stands for $\phi_1$, while $\phi'$ stands for
$\phi_3$ and the two fields with opposite charges are $\phi_1'$ and
$\phi_3'$. The fields with opposite charges to $A_+$ and
$A_-$ are $D^{(5)}_-$ and $D^{(5)}_+$, respectively, while the field with
opposite charges to $D_2$ is $D^{(3,4)}_{+-}$  and $\tilde{D}_2''$ stands
for $D^{(3,4)}_{-+}$, in the notation of the table \ref{matter1}. We did not include in table \ref{flat} the fields
$\tilde{\phi}_1$, $\phi_2$ and $\tilde{\phi}_3$, which have vanishing
charges. These fields are trivially flat directions in the
$U(1)_{1...5}$, but they are not flat under the anomalous $U(1)$.

For simplicity we rescaled the charges $Q_1$, $Q_3$ and $Q_A$ by a
factor 2 and the charges  $Q_2$, $Q_4$ and $Q_5$ by a factor 4.
The seventh column is given by

\be\hat{Q}=\frac{1}{18}(Q_A-Q_5+9\ Q_3)\ee
and, as explained in \cite{Cleaver:1998sm}, it will be useful for the
search of flat directions in the anomalous $U(1)$.

\beqn
 &\begin{tabular}{l|rrrrrrr}
 ~ & $\quad Q_1$ & $\quad Q_2$ & $\quad Q_3$ & $\quad Q_4$ &
$\quad Q_5$ & $\quad Q_A$ & $\quad \hat{Q}$\\
\hline
\hline
  $\phi^{(')}~ \surd^{(')}$  & 0 & 4& 0 & -4 & 4 & 4 & 0 \\
 $S_1^{(')},D_1$ & 1 & 2 & -1 & 0 & -12 & -3 & 0\\
$\tilde{S}_1^{(')}, \tilde{D}_1^{(')}$ & 1 & 2 & 1 & 0 & 4 & -5 & 0 \\
$S_2^{(')}, D_2 ~\surd$ & -1 & 4 & 0 & -2 & -2 & -2 & 0\\
$\tilde{S}_2^{(')}, \tilde{D}_2^{(')('')}$ & -1 & 0 & 0 & 2 & -6 & -6 & 0 \\
$S_3^{(')},D_3$ & 0 & 0 & 1 & -4 & -12 &-3 & 1\\
$\tilde{S}_3^{(')}, \tilde{D}_3^{(')}$ & 0 & 0 & -1 & -4 & 4 & -5 & -1\\
$N_1$ & -1 & 0 & -1 & -2 & -10 & -1 & 0\\
$N_2$ & 1 & -2 & 0 & 0 & -4 & -4 & 0\\
$N_3$ & 0 & 2 & -1 & 2 & 6 & -3 & -1\\
$A_+ ~\surd$ & -1 & 0 & 0 & -6 & 2 & 2 & 0\\
$A_- ~\surd$ &1 & 4 & 0 & 2 & 2 &2 & 0\\
$F^{(')}$ &1 & -1 & 2 & -1 &1 &1 & 1\\
$\tilde{F}^{(')}$ & -1 &1 & 0 &1 &15 & -3 & -1\\
$F_1$ & 0 & 3 &1 &1 & 11 & 2 & 0\\
$F_2$ & 0 & -1 & 1 & 5 & 7 & -2 & 0\\
$F_3$ & 0 & 1 &1 & -5 & 9 & 0 & 0\\
$F_4$ & 0 &  -3 & 1 &  -1 & 5 & -4 & 0\\
\end{tabular}
\label{flat}
\eeqn

As a first step we investigate the existence of flat directions involving
vacuum expectation values only for the fields which are singlets under
both the visible and the hidden gauge groups. These fields are
$\phi^{(')}\surd^{(')}$, $S_1^{(')}$,  $\tilde{S}_1^{(')}$, $S_2^{(')}$,
$\tilde{S}_2^{(')}$, $S_3^{(')}$, $\tilde{S}_3^{(')}$, $N_1$, $N_2$ and
$N_3$. Bearing in mind the equivalence in the charges for some fields,
these count as 11 fields and so, given the fact that we have to impose 5
constraints, the basis of flat directions should contain 6 elements. But
a simple Mathematica program can show that it is impossible to incorporate
the fields  $S_1^{(')}$, $S_3^{(')}$, $N_1$, $N_2$ and $N_3$ into the flat
directions. This leave us with 6 fields, so we expect a basis with just one
element. It turns out that, in respect with the charges of the remaining
fields, $Q_4$ and $Q_5$ are a linear combination of the previous $U(1)$'s,
so there are actually only 3 independent constraints and, hence, we
obtain three basis elements
\be \phi\bar{\phi}, ~~\bar{\phi}\tilde{S}_1^2\tilde{S}_2^2\tilde{S}_3^2,
{}~~\bar{\phi}^3\tilde{S}_1^2S_2^2\tilde{S}_3^2,\ee
where we expressed the flat directions as gauge invariant monomials.
For example, the monomial
$\bar{\phi}\tilde{S}_1^2\tilde{S}_2^2\tilde{S}_3^2$
corresponds to the following choice of VEVs

\be |\bar{\phi}|^2=|\psi|^2, ~ |\tilde{S}_1|^2=2|\psi|^2,
{}~ |\tilde{S}_2|^2=2|\psi|^2, ~ |\tilde{S}_3|^2=2|\psi|^2, \ee
for an arbitrary $|\psi|$.

Note that in the precedent basis any field A can be replaced with its
copy A$^{'}$. Any flat direction, $P$,  can be obtained from the elements of
the basis as

\be P^n = \prod_\alpha M_\alpha^{n_\alpha}, \ee
where $M_\alpha$ stand for the elements of the basis, $n$ is a positive
integer and $n_\alpha$ are integers \cite{Cleaver:1998sm}.

In order to obtain D-flat directions in the anomalous $U(1)$ we need to
construct invariant monomials containing the field $S_3^{(')}$, since this
is the only field with a positive $\hat{Q}$ charge $\footnote{The $\hat{Q}$
charge of an invariant monomial is equal, up to positive factors, with his
$Q_A$ charge, since the difference between the two is a linear combination
of $Q_{1...5}$, under which the invariant monomials have zero charge by
construction.}$, necessary to cancel the negative Fayet-Iliopoulos term
generated by the anomalous $U(1)$ $\footnote{In our model
${\rm Tr}~Q_A<0$.}$. And, since none of the elements of the basis contains
this field, we conclude that there are no flat directions involving only
VEVs of the singlets.

Therefore, we proceed with the analysis including also nonabelian fields
under the hidden gauge group.  This amounts to including all the fields in
table \ref{flat}, which contains 22 fields with non-equivalent charges.
Again, we look for a basis of gauge invariant monomials under $Q_{1...5}$.
Such a basis is given by
\ba &&\phi\bar{\phi}, ~~ D_2\bar{D_2}, ~~A_+\bar{A}_+, ~~A_-\bar{A}_-,
{}~~\bar{\phi}\tilde{S}_1^2\tilde{S}_2^2\tilde{S}_3^2,
{}~~\bar{\phi}^3\tilde{S}_1^2S_2^2\tilde{S}_3^2,~~\bar{\phi}A_+A_-,\nonumber\\
\nonumber\\
&&\bar{\phi}S_1^4N_1^2F^2\tilde{F}^4F_4^2,
{}~~\bar{\phi}S_1^2S_3^2N_3^2\tilde{F}^2F_4^2,
{}~~\bar{\phi}S_3^2N_1^2N_2^2N_3^4F^2\tilde{F}^2,\nonumber\\
\nonumber\\
&&\bar{\phi}S_3^2N_1^2N_2^2N_3^4F_1^2F_4^2,
{}~~\bar{\phi}S_3^2N_1^2N_2^2N_3^4F_2^2F_3^2,
{}~~S_1^2\tilde{S}_2S_3\tilde{S_3}\bar{A}_+\tilde{F}^2F_4^2,\nonumber\\
\nonumber\\
&&S_1^3\tilde{S}_2^3S_3\tilde{S_3}^2N_1N_2\bar{A}_+^3\tilde{F}^3F_3^2F_4^3,
{}~~\tilde{S}_2^5S_3\tilde{S_3}^5\bar{A}_+^5F_3^3F_4,
{}~~\bar{\phi}S_1^{10}\tilde{S}_2^2S_3^2\tilde{F}^8F_4^8,\nonumber\\
\nonumber\\
&&S_1^9\tilde{S}_2^2S_3^2N_1N_2\bar{A}_+\tilde{F}^8F_4^8
\label{nabasis}
\ea
where, again, any field can be replaced with one of its copies with equal
$Q_{1...5}$ charges. All the elements of the basis have negative or vanishing
$\hat{Q}$ charges, but, since some of the elements contain the fields $S_3$ and
$F$, which have positive
$\hat{Q}$ charge, and, since flat directions can be obtained as a combination
of the basis elements with negative powers, we cannot conclude immediately that
there are no D-flat directions under the anomalous $U(1)$.  Nevertheless, a
simple Mathematica program shows that it is impossible to obtain viable
invariant monomials with positive $\hat{Q}$ charge, by viable meaning that
the fields that do not have a partner field with opposite charges should appear
with positive powers in the monomials. We conclude that there are no flat
directions involving only singlets of the visible gauge group.

Therefore, the only possibility
to obtain flat directions which do not break
electric charge
is to consider the option of giving a VEV also to
the neutral component of the Higgs field, in which case the flat directions
would break the electroweak symmetry.
The Higgs doublets in our model have the following charges:
\beqn
 &\begin{tabular}{l|rrrrrrr}
 ~ & $\quad Q_1$ & $\quad Q_2$ & $\quad Q_3$ & $\quad Q_4$ & $\quad Q_5$ &
$\quad Q_A$ & $\quad \hat{Q}$\\
\hline
\hline
  $h~ \surd$  & 0 & 4& 0 & 4 & -4 & -4 & 0
\end{tabular}
\eeqn
and including them into our analysis amounts to adding the invariant monomial
$\bar{\phi}hS_1^2S_3\tilde{S}_3\tilde{F}^2F_4^2$ to the basis \ref{nabasis}.
The new basis element also has a negative $\hat{Q}$ charge and, again, it turns
out to be impossible to construct flat directions with positive  $\hat{Q}$
charge. This means that the only stable vacuum solutions of our model are the
ones that break the Standard Model gauge group. We did not search for such flat
directions (i.e. including in the analysis all the states in the spectrum), but preliminary 
analysis indicates their existence.

\section{Conclusions}\label{conclude}
The quasi--realistic free fermionic models,
which are related to $Z_2\times Z_2$ orbifold compactifications at special points
in the moduli space \cite{foc},
produced some of the most realistic string models
constructed to date. The underlying $Z_2\times Z_2$ orbifold structure
dictates that the models generically contain three pairs of untwisted
Higgs multiplets. The reduction of the Higgs states
to the Standard Model spectrum relied in the past
on the analysis of supersymmetric flat directions, which
give superheavy mass to the dispensable Higgs respresentations.
In this paper we investigated the possibility of removing the extra
Higgs multiplets by using the free fermion boundary conditions
and directly at the string level, rather than in the effective low energy 
field theory.

The mechanism presented relies on the possibility of assigning boundary
conditions that are asymmetric between the left-- and right--moving
internal fermions. In this respect, it should be noted that all
$Z_2\times Z_2$ orbifold models constructed to date use symmetric
boundary conditions and is therefore of immense interest to incorporate
asymmetric twistings in these bosonic constructions.

We also incorporated the Higgs reduction mechanism in a three
generation model and therefore obtained for the first time a model
with three generations of chiral fermions, arising from the $16$ spinorial
representation of $SO(10)$, and one pair of electroweak Higgs doublet,
arising from the $10$ vectorial representation of $SO(10)$, directly
at the string level.

An additional effect of the asymmetric twisting is the reduction of the moduli
space. In the past this has been demonstrated with respect to the
untwisted moduli that parametrise the shape and size of the dimensional
compactified manifold \cite{moduli}.
The effect of the asymmetric boundary conditions that result in
untwisted Higgs reduction is, additionally, to reduce the untwisted Standard Model
singlet spectrum. The consequence is that the supersymmetric
moduli space is more restrictive. In fact, we demonstrated that in the model
presented here there do not exist flat directions that preserve the Standard Model
gauge group. This is in fact a welcomed situation, as it is likely that in
the vast space of three generation free fermionic models \cite{fknr}
there do exist vacua that accommodate a viable spectrum with the
Higgs reduction mechanism articulated here, and supersymmetric
flat directions. In such models, however, the supersymmetric moduli
space will be much reduced, hence increasing their predictive power.

\bigskip
\medskip
\leftline{\large\bf Acknowledgments}
\medskip

AEF would like to thank the CERN and Oxford theory departments for
hospitality. CT would like to thank Matteo Cardella and Thomas Mohaupt for useful discussions.
This work was supported by the PPARC and the University of Liverpool.



\bigskip
\medskip

\bibliographystyle{unsrt}

\begin{thebibliography}{99}
\bibitem{gutsreviews} For reviews and references see {\it e.g.}:\\
		      P.\ Langacker, \PRP{72}{1981}{185};\\
C.~Kounnas, A.~Masiero, D.V.~Nanopoulos and K.A.~Olive,
Grand Unification With And Without Supersymmetry And Cosmological
Implications (World Scientific, Singapore, 1984).

\bibitem{hete} D.J.\ Gross, J.A.\ Harvey, J.A.\ Martinec and R.\ Rohm,
		\NPB{256}{1986}{253};\\
	       P.\ Candelas, G.T.\ Horowitz, A.\ Strominger and E.\ Witten,
			\NPB{258}{1985}{46}.
\bibitem{resurgence} See {\it e.g.}:
J.~Giedt, Annals Phys.\  {\bf 297} (2002) 67;\\
S.~Forste, H.P.~Nilles, P.K.S.~Vaudrevange and A.~Wingerter,
						\PRD{70}{2004}{106008};\\
T. Kobayashi, S. Raby and R.J. Zhang, \NPB{704}{2005}{3};\\
V. Bouchard and R. Donagi, \PLB{633}{2006}{783};\\
V. Braun, Y.H. He, B.A. Ovrut and T. Pantev, \JHEP{0605}{2006}{043};\\
R. Blumenhagen, S. Moster and T. Weigand, \NPB{751}{2006}{186};\\
W.~Buchmuller, K.~Hamaguchi, O.~Lebedev and M.~Ratz, \PRL{96}{2006}{121602};\\
J.~E.~Kim and B.~Kyae, arXiv:hep-th/0608085;\\
J.~E.~Kim and B.~Kyae, arXiv:hep-th/0608086.



\bibitem{fsu5} I.\ Antoniadis, J.\ Ellis, J.\ Hagelin and D.V.\ Nanopoulos
                \PLB{231}{1989}{65};\\
		J.L. Lopez, D.V. Nanopoulos and K. Yuan, \NPB{399}{1993}{3}.

\bibitem{fny} A.E.\ Faraggi, D.V.\ Nanopoulos and K.\ Yuan,
                                                 \NPB{335}{1990}{347}.

\bibitem{fc} \AEF, \PRD{46}{1992}{3204}.

\bibitem{alr} I.\ Antoniadis, G.K.\ Leontaris and J.\ Rizos,
				\PLB{245}{1990}{161};\\
		G.K.\ Leontaris and J.\ Rizos,
				\NPB{554}{1999}{3};\\

\bibitem{eu}A.E.\ Faraggi, \PLB{278}{1992}{131};
        		\NPB{387}{1992}{239}; \NPB{403}{1993}{101}.

\bibitem{top} \AEF, \PLB{274}{1992}{47}; \PRD{47}{1993}{5021};
		\PLB{377}{1996}{43};
			\NPB{487}{1997}{55}.

\bibitem{nahe} \AEF~and D.V.\ Nanopoulos, \PRD{48}{1993}{3288};\\
               \AEF, \NPB{387}{1992}{239}.

\bibitem{cfn} G.B.\ Cleaver, \AEF~ and D.V.\ Nanopoulos,
                       \PLB{455}{1999}{135}; \IJMP{16}{2001}{425};\\
              G.B.\ Cleaver, \AEF, D.V.\ Nanopoulos and J.W.\ Walker,
                                        \NPB{593}{2001}{471};
                                        \NPB{620}{2002}{259}.

\bibitem{cfs} G.B.\ Cleaver, A.E.\ Faraggi and C.\ Savage,
                                \PRD{63}{2001}{066001};\\
              G.B.\ Cleaver, D.J.\ Clements and A.E.\ Faraggi,
                        \PRD{65}{2002}{106003}.

\bibitem{fff}  H.\ Kawai, D.C.\ Lewellen, and S.H.-H.\ Tye,
					\NPB{288}{1987}{1};\\
               I.\ Antoniadis, C.\ Bachas, and C.\ Kounnas,
	       \NPB{289}{1987}{87};\\
	       I.\ Antoniadis and C.\ Bachas, \NPB{289}{1987}{87}.

\bibitem{kln} S. Kalara, J.L. Lopez and D.V. Nanopoulos, \NPB{353}{1991}{650}.

\bibitem{costas} S.\ Ferrara, L.\ Girardello, C.\ Kounnas and M.\ Porrati,
                 \PLB{194}{1987}{368};\\
		 S.\ Ferrara, C.\ Kounnas, M.\ Porrati and F.\ Zwirner,
		 \PLB{194}{1987}{366}.

\bibitem{cfnooij} G.B.\ Cleaver, A.E.\ Faraggi and S.E.M.\ Nooij,
                                \NPB{672}{2003}{64}.

\bibitem{dsw}M. Dine, N. Seiberg and E. Witten, \NPB{289}{1987}{585}.

\bibitem{ps} \AEF, \NPB{428}{1994}{111}; \PLB{520}{2001}{337}.

\bibitem{ffl2} \AEF, \IJMP{14}{1999}{1663}.

\bibitem{Cleaver:1998sm}
G.~Cleaver, M.~Cvetic, J.~R.~Espinosa, L.~L.~Everett, P.~Langacker and J.~Wang,
  Phys.\ Rev.\ D {\bf 59} (1999) 115003

\bibitem{foc} A.E.\ Faraggi, \PLB{326}{1994}{62}; hep-th/9511093;\\
              J.\ Ellis, A.E.\ Faraggi and D.V.\ Nanopoulos,
                                                \PLB{419}{1998}{123};\\
         P.\ Berglund, J.\ Ellis, A.E.\ Faraggi, D.V.\ Nanopoulos and
                                                Z.\ Qiu,
                        \PLB{433}{1998}{269}; \IJMP{15}{2000}{1345};\\
         A.E. Faraggi, \PLB{544}{2002}{207}; hep-th/0411118;\\
         \AEF~and R. Donagi, \NPB{694}{2004}{187};\\
	\AEF, S.\ F\"orste, M.C.\ Timirgaziu, \JHEP{0608}{2006}{057}.

\bibitem{moduli} \AEF, \NPB{728}{2005}{83}.

\bibitem{fknr} \AEF, C. Kounnas, S. Nooij and J. Rizos,
               hep-th/0311058; \NPB{695}{2004}{41}; hep-th/0606144.

\end{thebibliography}

\vfill\eject
\newpage
\renewcommand{\baselinestretch}{1.3}
\begin{table}
{\bf Table \ref{matter1}}
\begin{eqnarray*}
\begin{tabular}{|c|c|c|rrrrrrrr|c|rr|}
\hline
  $F$ & SEC & $SU(3)\times$&$Q_{C}$ & $Q_L$ & $Q_1$ &
   $Q_2$ & $Q_3$ & $Q_{4}$ & $Q_{5}$ & $Q_6$ &
   $SU(2)_{1,..,6}^6$ & $Q_7$ & $Q_{8}$\\
   $$ & $$ & $SU(2)$ & $$ & $$ & $$ & $$ & $$ & $$ & $$ & $$ & $$  & $$ & $$ \\
\hline
   $L_1$ & $b_1$      & $(1,2)$ & $-{3\over2}$ & $0$ &
   $-{1\over2}$ & $0$ & $0$ & $-{1\over2}$ & $0$ & $0$ &
   $(1,1,1,1,1,1)$ & $0$ & $0$ \\
   $Q_1$ &            & $(3,2)$ & $ {1\over2}$ & $0$ &
   $-{1\over2}$ & $0$ & $0$ & $1\over2$ & $0$ & $0$ &
   $(1,1,1,1,1,1)$ & $0$ & $0$ \\
   ${d}_{L_{1}}^c$ &            & $({\bar 3},1)$ & $-{1\over2}$ & $1$ &
   $-{1\over2}$ & $0$ & $0$ & $-{1\over2}$ & $0$ & $0$ &
   $(1,1,1,1,1,1)$ & $0$ & $0$ \\
   ${N}_{L_1}^c $&            & $(1,1)$ & ${3\over2}$ & $-1$ &
   $-{1\over 2}$ & $0$ & $0$ & $-{1\over2}$ & $0$ & $0$ &
   $(1,1,1,1,1,1)$ & $0$ & $0$ \\
   ${u}_{L_1}^c$ &            & $({\bar 3},1)$ & $-{1\over2}$ & $-1$ &
   $-{1\over2}$ & $0$ & $0$ & $1\over2$ & $0$ & $0$ &
   $(1,1,1,1,1,1)$ & $0$ & $0$ \\
   ${e}_{L_1}^c$ &            & $(1,1)$ & ${3\over2}$ & $1$ &
   $-{1\over 2}$ & $0$ & $0$ & $1\over2$ & $0$ & $0$ &
   $(1,1,1,1,1,1)$ & $0$ & $0$ \\
\hline
   $L_2$ & $b_2$      & $(1,2)$ & $-{3\over2}$ & $0$ &
   $0$ & $-{1\over2}$ & $0$ & $0$ & ${1\over2}$ & $0$ &
   $(1,1,1,1,1,1)$ & $0$ & $0$ \\
   $Q_2$ &            & $(3,2)$ & $ {1\over2}$ & $0$ &
   $0$ & $-{1\over2}$ & $0$ & $0$ & $-{1\over2}$ & $0$ &
   $(1,1,1,1,1,1)$ & $0$ & $0$ \\
   $d_{L_2}^c$ &            & $({\bar 3},1)$ & $-{1\over2}$ & $1$ &
   $0$ & $-{1\over2}$ & $0$ & $0$ & $-{1\over2}$ & $0$ &
   $(1,1,1,1,1,1)$ & $0$ & $0$ \\
   ${N}_{L_2}^c$ &            & $(1,1)$ & ${3\over2}$ & $-1$ &
   $0$ & $-{1\over 2}$ & $0$ & $0$ & $-{1\over2}$ & $0$ &
   $(1,1,1,1,1,1)$ & $0$ & $0$ \\
   $u_{L_2}^c$ &            & $({\bar 3},1)$ & $-{1\over2}$ & $-1$ &
   $0$ & $-{1\over2}$ & $0$ & $0$ & $1\over2$ & $0$ &
   $(1,1,1,1,1,1)$ & $0$ & $0$ \\
   ${e}_{L_2}^c$ &            & $(1,1)$ & ${3\over2}$ & $1$ &
   $0$ & $-{1\over 2}$ & $0$ & $0$ & $1\over2$ & $0$ &
   $(1,1,1,1,1,1)$ & $0$ & $0$ \\
\hline
   $L_3$ & $b_3$      & $(1,2)$ & $-{3\over2}$ & $0$ &
   $0$ & $0$ & ${1\over2}$ & $0$ & $0$ & $1\over2$ &
   $(1,1,1,1,1,1)$ & $0$ & $0$ \\
   $Q_3$ &            & $(3,2)$ & $ {1\over2}$ & $0$ &
   $0$ & $0$ & ${1\over2}$ & $0$ & $0$ & $-{1\over2}$ &
   $(1,1,1,1,1,1)$ & $0$ & $0$ \\
   $d_{L_3}^c$ &            & $({\bar 3},1)$ & $-{1\over2}$ & $1$ &
   $0$ & $0$ & ${1\over2}$ & $0$ & $0$ & $-{1\over2}$ &
   $(1,1,1,1,1,1)$ & $0$ & $0$ \\
   ${N}_{L_3}^c$ &            & $(1,1)$ & ${3\over2}$ & $-1$ &
   $0$ & $0$ & ${1\over 2}$ & $0$ & $0$ & $-{1\over2}$ &
   $(1,1,1,1,1,1)$ & $0$ & $0$ \\
   $u_{L_3}^c$ &            & $({\bar 3},1)$ & $-{1\over2}$ & $-1$ &
   $0$ & $0$ & ${1\over2}$ & $0$ & $0$ & $1\over2$ &
   $(1,1,1,1,1,1)$ & $0$ & $0$ \\
   ${e}_{L_3}^c$ &            & $(1,1)$ & ${3\over2}$ & $1$ &
   $0$ & $0$ & ${1\over 2}$ & $0$ & $0$ & $1\over2$ &
   $(1,1,1,1,1,1)$ & $0$ & $0$ \\
\hline
   $ h$ & ${\rm NS}$ & $(1,2)$ & $0$ & $-1$ &
   $0$ & $0$ & $1$ & $0$ & $0$ & $0$ &
   $(1,1,1,1,1,1)$ & $0$ & $0$ \\
   $\bar h$ & $$ & $(1,2)$ & $0$ & $1$ &
   $0$ & $0$ & $-1$ & $0$ & $0$ & $0$ &
   $(1,1,1,1,1,1)$ & $0$ & $0$ \\
   $\phi_{1}$ &                       & $(1,1)$ & $0$ & $0$ &
   $0$ & $0$ & $0$ & $0$ & $1$ & $0$ &
   $(1,1,1,1,1,1)$ & $0$ & $0$ \\
   $\phi_{1}'$ &                       & $(1,1)$ & $0$ & $0$ &
   $0$ & $0$ & $0$ & $0$ & $-1$ & $0$ &
   $(1,1,1,1,1,1)$ & $0$ & $0$ \\
   $\tilde\phi_{1}$ &                 & $(1,1)$ & $0$ & $0$ &
   $0$ & $0$ & $0$ & $0$ & $0$ & $0$ &
   $(1,1,1,1,1,1)$ & $0$ & $0$ \\
   $\phi_{2}$ &                 & $(1,1)$ & $0$ & $0$ &
   $0$ & $0$ & $0$ & $0$ & $0$ & $0$ &
   $(1,1,1,1,1,1)$ & $0$ & $0$ \\
   $\phi_{3}$ &                 & $(1,1)$ & $0$ & $0$ &
   $0$ & $0$ & $0$ & $0$ & $1$ & $0$ &
   $(1,1,1,1,1,1)$ & $0$ & $0$ \\
   $\phi_{3}'$ &                 & $(1,1)$ & $0$ & $0$ &
   $0$ & $0$ & $0$ & $0$ & $-1$ & $0$ &
   $(1,1,1,1,1,1)$ & $0$ & $0$ \\
   $\tilde\phi_{3}$ &                 & $(1,1)$ & $0$ & $0$ &
   $0$ & $0$ & $0$ & $0$ & $0$ & $0$ &
   $(1,1,1,1,1,1)$ & $0$ & $0$ \\
   \hline
\end{tabular}
\label{matter1}
\end{eqnarray*}

\end{table}

\vfill
\eject

\begin{table}
\begin{eqnarray*}
\begin{tabular}{|c|c|c|rrrrrrrr|c|rr|}
\hline
  $F$ & SEC & $SU(3)\times $&$Q_{C}$ & $Q_L$ & $Q_1$ &
   $Q_2$ & $Q_3$ & $Q_{4}$ & $Q_{5}$ & $Q_6$ &
   $SU(2)_{1,..,6}^6$ & $Q_{7}$ & $Q_{8}$\\
   $$ & $$ & $SU(2)$ & $$ & $$ & $$ & $$ & $$ & $$ & $$ & $$ & $$  & $$ & $$ \\
\hline
$C^{-+}_+$ & $1+b_4$ & $(1,1)$ & $0$ & $-1$ &
   $-{1\over2}$ & $0$ & ${1\over2}$ & $0$ & $1\over2$ & $0$ &
   $(1,1,1,1,1,1)$ & $1$ & $0$ \\
   $C^{-+}_-$ & $+\beta+2\gamma$ & $(1,1)$ & $0$ & $1$ &
   $-{1\over2}$ & $0$ & ${1\over2}$ & $0$ & $1\over2$ & $0$ &
   $(1,1,1,1,1,1)$ & $-1$ & $0$ \\
   $D_+$ &    & $(1,2)$ & $0$ & $0$ &
   $-{1\over2}$ & $0$ & $-{1\over2}$ & $0$ & $-{1\over2}$ & $0$ &
   $(1,1,1,1,1,1)$ & $1$ & $0$ \\
   $D_-$ &              & $(1,2)$ & $0$ & $0$ &
   ${1\over2}$ & $0$ & ${1\over2}$ & $0$ & $-{1\over2}$ & $0$ &
   $(1,1,1,1,1,1)$ & $-1$ & $0$ \\
   $C^{+-}_+$ &              & $(1,1)$ & $0$ & $-1$ &
   ${1\over2}$ & $0$ & $-{1\over2}$ & $0$ & $1\over2$ & $0$ &
   $(1,1,1,1,1,1)$ & $1$ & $0$ \\
   $C^{+-}_-$ &              & $(1,1)$ & $0$ & $1$ &
   ${1\over2}$ & $0$ & $-{1\over2}$ & $0$ & $1\over2$ & $0$ &
   $(1,1,1,1,1,1)$ & $-1$ & $0$ \\
\hline
   $T_+$ & $1+b_4$ & $(\bar{3},1)$ & $-{1\over2}$ & $0$ & 
$0$ & $-{1\over2}$ &
   $0$ & $0$ & $-{1\over2}$ & $0$ & $(1,1,1,1,1,1)$ & $0$ &
   $1$ \\
$C_-$ & $+\beta$ & $(1,1)$ & ${3\over2}$ & $0$ & $0$ & $-{1\over2}$ &
   $0$ & $0$ & $-{1\over2}$ & $0$ & $(1,1,1,1,1,1)$ & $0$ &
   $-1$ \\
$C_+$ & $$ & $(1,1)$ & $-{3\over2}$ & $0$ & $0$ & ${1\over 2}$ &
   $0$ & $0$ & $-{1\over2}$ & $0$ & $(1,1,1,1,1,1)$ & $0$ &
   $1$ \\
$T_-$ & $$ & $(3,1)$ & ${1\over2}$ & $0$ & $0$ & ${1\over 2}$ &
   $0$ & $0$ & $-{1\over2}$ & $0$ & $(1,1,1,1,1,1)$ & $0$ &
   $-1$ \\
\hline
   $D_1$ & $b_1+2\gamma$ & $(1,1)$ & $0$ & $0$ & $0$ &
    $-{1\over2}$ & ${1\over 2}$ & $-{1\over2}$ & $0$ & $0$
& $(1,1,2,1,1,2)$ & $0$ &
   $0$ \\
   $S_1$ & $$      & $(1,1)$ & $0$ & $0$ & $0$ &
    $-{1\over2}$ & ${1\over 2}$ & $-{1\over2}$ & $0$ & $0$ & $(1,1,1,1,1,1)$ &
    $-1$ &
    $-1$ \\
   $S_1'$ &  		     & $(1,1)$ & $0$ & $0$ & $0$ &
    $-{1\over2}$ & ${1\over 2}$ & $-{1\over2}$ & $0$ & $0$ & $(1,1,1,1,1,1)$
& $1$ &
   $1$ \\
   $\tilde{S}_1$ &  	     & $(1,1)$ & $0$ & $0$ & $0$ &
    $-{1\over2}$ & ${1\over 2}$ & $1\over2$ & $0$ & $0$ & $(1,1,1,1,1,1)$ &
    $-1$ &
    $1$ \\
$\tilde{S}_1'$ &  	     & $(1,1)$ & $0$ & $0$ & $0$ &
    $-{1\over2}$ & ${1\over 2}$ & $1\over2$ & $0$ & $0$ & $(1,1,1,1,1,1)$ &
    $1$ &
    $-1$ \\
\hline
   $S_2$ & $b_2+2\gamma$ & $(1,1)$ & $0$ & $0$ & $-{1\over 2}$ &
   $0$ & ${1\over2}$ & $0$ & $1\over2$ & $0$ & $(1,1,1,1,1,1)$ & $-1$ &
   $1$ \\
   $S_2'$ & $$ & $(1,1)$ & $0$ & $0$ & $-{1\over2}$ &
   $0$ & ${1\over2}$ & $0$ & $1\over2$ & $0$ & $(1,1,1,1,1,1)$ & $1$ &
   $-1$ \\
   $D_2$ &  		     & $(1,1)$ & $0$ & $0$ & $-{1\over2}$ &
   $0$ & ${1\over2}$ & $0$ & $1\over2$ & $0$ & $(1,1,2,1,1,2)$ & $0$ &
   $0$ \\
   $\tilde{S}_2$ &  	     & $(1,1)$ & $0$ & $0$ & $-{1\over 2}$ &
   $0$ & ${1\over2}$ & $0$ & $-{1\over2}$ & $0$ & $(1,1,1,1,1,1)$ &
   $-1$ &
   $-1$  \\
$\tilde{S}_2'$ &  	     & $(1,1)$ & $0$ & $0$ & $-{1\over 2}$ &
   $0$ & ${1\over2}$ & $0$ & $-{1\over2}$ & $0$ & $(1,1,1,1,1,1)$ &
   $1$ &
   $1$  \\
\hline
  $S_3$ & $b_3+2\gamma$ & $(1,1)$ & $0$ & $0$ & $-{1\over 2}$ &
   $-{1\over2}$ & $0$ & $0$ & $0$ & $1\over2$ & $(1,1,1,1,1,1)$ & $-1$ &
   $1$ \\
 $S_3'$ & $$ & $(1,1)$ & $0$ & $0$ & $-{1\over 2}$ &
   $-{1\over2}$ & $0$ & $0$ & $0$ & $1\over2$ & $(1,1,1,1,1,1)$ & $1$ &
   $-1$ \\
$\tilde{S_3}$ & $$ & $(1,1)$ & $0$ & $0$ & $-{1\over 2}$ &
   $-{1\over2}$ & $0$ & $0$ & $0$ & $-{1\over2}$ & $(1,1,1,1,1,1)$ & $-1$ &
   $-1$ \\
$\tilde{S}_3'$ & $$ & $(1,1)$ & $0$ & $0$ & $-{1\over 2}$ &
   $-{1\over2}$ & $0$ & $0$ & $0$ & $-{1\over2}$ & $(1,1,1,1,1,1)$ & $1$ &
   $1$ \\
$\tilde{D}_3$ & $$ & $(1,1)$ & $0$ & $0$ & $-{1\over 2}$ &
   $-{1\over2}$ & $0$ & $0$ & $0$ & $-{1\over2}$ & $(1,1,2,1,1,2)$ & $0$ &
   $0$ \\
\hline
   $A_+$ & $b_4+2\gamma$     & $(1,1)$ & $0$ & $0$ &
   $-{1\over2}$ & $0$ & $-{1\over 2}$ &
   $0$ & $1\over2$ & $0$ &
   (2,1,1,1,1,1) & $0$ & $1$   \\
 $A_-$ & $$     & $(1,1)$ & $0$ & $0$ &
   ${1\over2}$ & $0$ & ${1\over 2}$ &
   $0$ & $1\over2$ & $0$ &
   (2,1,1,1,1,1) & $0$ & $-1$   \\
 \hline
\end{tabular}
\label{matter2}
\end{eqnarray*}

\end{table}

\vfill
\eject

\begin{table}
\begin{eqnarray*}
\begin{tabular}{|c|c|c|rrrrrrrr|c|rr|}
\hline
  $F$ & SEC & $SU(3)\times $&$Q_{C}$ & $Q_L$ & $Q_1$ &
   $Q_2$ & $Q_3$ & $Q_{4}$ & $Q_{5}$ & $Q_6$ &
   $ SU(2)_{1,..,6}^6$ & $Q_{7}$ & $Q_{8}$\\
   $$ & $$ & $SU(2)$ & $$ & $$ & $$ & $$ & $$ & $$ & $$ & $$ & $$  & $$ & $$ \\
\hline
$\tilde{D}_1$ & $1+b_2+$ & (1,1) & $0$  & $0$ &
   $0$ & $-{1\over2}$ & ${1\over2}$ &
   $1\over2$ & $0$ & $0$ &
   (1,2,1,2,1,1) & $0$ & $0$  \\
   $\tilde{D}_1'$ & $b_3+2\gamma$ & (1,1) & $0$  & $0$ &
   $0$ & $-{1\over2}$ & ${1\over2}$ &
   $1\over2$ & $0$ & $0$ &
   (2,1,1,1,2,1) & $0$ & $0$  \\
\hline
 $\tilde{D}_2$ & $1+b_1+$ & (1,1) & $0$  & $0$ &
   $-{1\over2}$ & $0$ & ${1\over2}$ &
   $0$ & $-{1\over2}$ & $0$ &
   (2,1,1,1,2,1) & $0$ & $0$  \\
   $\tilde{D}_2'$ & $b_3+2\gamma$ & (1,1) & $0$  & $0$ &
   $-{1\over2}$ & $0$ & ${1\over2}$ &
   $0$ & $-{1\over2}$ & $0$ &
   (1,2,1,2,1,1) & $0$ & $0$  \\
\hline
 $$ & $1+b_1+b_2$ & (1,1) & $3\over4$  & $1\over2$ &
   $1\over4$ & $-{1\over4}$ & $-{1\over4}$ &
   $1\over2$ & $0$ & $-{1\over2}$ &
   (1,1,1,2,1,1) & $1\over2$ & $-{1\over2}$  \\
 $$ & $b_4\pm\gamma$ & (1,1) & $-{3\over4}$  & $-{1\over2}$ &
   $-{1\over4}$ & $1\over4$ & $1\over4$ &
   $-{1\over2}$ & $0$ & $-{1\over2}$ &
   (1,1,1,2,1,1) & $-{1\over2}$ & $1\over2$  \\
\hline
 $\tilde{D}_3'$ & $1+b_1+$ & (1,1) & $0$  & $0$ &
   $-{1\over2}$ & $-{1\over2}$ & $0$ &
   $0$ & $0$ & $-{1\over2}$ &
   (2,1,1,1,2,1) & $0$ & $0$  \\
   $D_3$ & $b_2+2\gamma$ & (1,1) & $0$  & $0$ &
   $-{1\over2}$ & $-{1\over2}$ & $0$ &
   $0$ & $0$ & $1\over2$ &
   (1,2,1,2,1,1) & $0$ & $0$  \\
\hline
 $$ & $1+b_1+$ & (1,1) & $0$  & $-1$ &
   $0$ & $0$ & $0$ &
   $-{1\over2}$ & $-{1\over2}$ & ${1\over2}$ &
   (2,1,1,1,1,1) & $0$ & $0$  \\
 $$ & $b_2+b_3+$ & (1,1) & $0$  & $1$ &
   $0$ & $0$ & $0$ &
   ${1\over2}$ & $-{1\over2}$ & ${1\over2}$ &
   (2,1,1,1,1,1) & $0$ & $0$  \\
   $$ & $\beta+2\gamma$ & (1,1) & $0$  & $1$ &
   $0$ & $0$ & $0$ &
   $-{1\over2}$ & $-{1\over2}$ & $-{1\over2}$ &
   (2,1,1,1,1,1) & $0$ & $0$  \\
   $$ & $$ & (1,1) & $0$  & $-1$ &
   $0$ & $0$ & $0$ &
   ${1\over2}$ & $-{1\over2}$ & $-{1\over2}$ &
   (2,1,1,1,1,1) & $0$ & $0$  \\
\hline
 $D^{(3,4)}_{-+}$ & $1+b_1+$ & (1,1) & $0$  & $0$ &
   $-{1\over2}$ & $0$ & ${1\over2}$ &
   $0$ & $-{1\over2}$ & $0$ &
   (1,1,2,2,1,1) & $0$ & $0$  \\
 $D^{(5)}_{+}$ & $b_2+b_3$ & (1,1) & $0$  & $0$ &
    $-{1\over2}$ & $0$ & $-{1\over2}$ &
   $0$ & $-{1\over2}$ & $0$ &
   (1,1,1,1,2,1) & $1$ & $0$  \\
 $D^{(5)}_{-}$ & $\alpha+2\gamma$ & (1,1) & $0$  & $0$ &
    ${1\over2}$ & $0$ & ${1\over2}$ &
   $0$ & $-{1\over2}$ & $0$ &
   (1,1,1,1,2,1) & $-1$ & $0$  \\
$D^{(3,4)}_{+-}$ & $$ & (1,1) & $0$  & $0$ &
    ${1\over2}$ & $0$ & $-{1\over2}$ &
   $0$ & $-{1\over2}$ & $0$ &
   (1,1,2,2,1,1) & $0$ & $0$  \\
\hline
$D^{(6)}_{+-}$ & $\pm\gamma$ & (1,1) & $3\over4$  & $1\over2$ &
   ${1\over4}$ & $1\over4$ & $1\over4$ &
   ${1\over2}$ & $-{1\over2}$ & $0$ &
   (1,1,1,1,1,2) & $1\over2$ & $-{1\over2}$  \\
     $D^{(6)}_{--}$ & $$ & (1,1) & $3\over4$  & $1\over2$ &
   $1\over4$ & $1\over4$ & $1\over4$ &
   $-{1\over2}$ & $-{1\over2}$ & $0$ &
   (1,1,1,1,1,2) & $1\over2$ & $-{1\over2}$  \\
   $D^{(6)}_{++}$ & $$ & (1,1) & $-{3\over4}$  & $-{1\over2}$ &
   $-{1\over4}$ & $-{1\over4}$ & $-{1\over4}$ &
   ${1\over2}$ & $-{1\over2}$ & $0$ &
   (1,1,1,1,1,2) & $-{1\over2}$ & ${1\over2}$  \\
   $D^{(6)}_{-+}$ & $$ & (1,1) & $-{3\over4}$  & $-{1\over2}$ &
   $-{1\over4}$ & $-{1\over4}$ & $-{1\over4}$ &
   $-{1\over2}$ & $-{1\over2}$ & $0$ &
   (1,1,1,1,1,2) & $-{1\over2}$ & ${1\over2}$  \\
    \hline
$D^{(3)}_{--}$ & $b_1+b_3$ & (1,1) & $3\over4$  & $1\over2$ &
   $-{1\over4}$ & $1\over4$ & $-{1\over4}$ &
   $0$ & $-{1\over2}$ & $-{1\over2}$ &
   (1,1,2,1,1,1) & $-{1\over2}$ & $1\over2$  \\
   $D^{(3)}_{+-}$ & $\pm\gamma$ & (1,1) & $3\over4$  & $1\over2$ &
   $-{1\over4}$ & $1\over4$ & $-{1\over4}$ &
   $0$ & $-{1\over2}$ & $1\over2$ &
   (1,1,2,1,1,1) & $-{1\over2}$ & $1\over2$  \\
   $D^{(3)}_{-+}$ & $$ & (1,1) & $-{3\over4}$  & $-{1\over2}$ &
   ${1\over4}$ & $-{1\over4}$ & ${1\over4}$ &
   $0$ & $-{1\over2}$ & $-{1\over2}$ &
   (1,1,2,1,1,1) & ${1\over2}$ & $-{1\over2}$  \\
   $D^{(3)}_{++}$ & $$ & (1,1) & $-{3\over4}$  & $-{1\over2}$ &
   ${1\over4}$ & $-{1\over4}$ & ${1\over4}$ &
   $0$ & $-{1\over2}$ & $1\over2$ &
   (1,1,2,1,1,1) & ${1\over2}$ & $-{1\over2}$  \\
  \hline
$F$ & $1+b_3+\alpha$ & (1,1) & $3\over4$  & $-{1\over2}$ &
   $1\over4$ & $-{1\over4}$ & $-{1\over4}$ &
   $1\over2$ & $0$ & ${1\over2}$ &
   (1,1,2,1,1,1) & $-{1\over2}$ & $-{1\over2}$  \\
$F'$ & $\beta\pm\gamma$ & (1,1) & $3\over4$  & $-{1\over2}$ &
   $1\over4$ & $-{1\over4}$ & $-{1\over4}$ &
   $1\over2$ & $0$ & ${1\over2}$ &
   (1,1,1,1,1,2) & $1\over2$ & ${1\over2}$  \\
$\tilde{F}$ & $$ & (1,1) & $-{3\over4}$  & ${1\over2}$ &
   $-{1\over4}$ & ${1\over4}$ & ${1\over4}$ &
   $1\over2$ & $0$ & $-{1\over2}$ &
   (1,1,2,1,1,1) & $1\over2$ & ${1\over2}$  \\
$\tilde{F}'$ & $$ & (1,1) & $-{3\over4}$  & ${1\over2}$ &
   $-{1\over4}$ & ${1\over4}$ & ${1\over4}$ &
   $1\over2$ & $0$ & $-{1\over2}$ &
   (1,1,1,1,1,2) & $-{1\over2}$ & $-{1\over2}$  \\
\hline
\end{tabular}
\label{matter32}
\end{eqnarray*}

\end{table}

\vfill
\eject

\begin{table}
\begin{eqnarray*}
\begin{tabular}{|c|c|c|rrrrrrrr|c|rr|}
\hline
  $F$ & SEC & $SU(3)\times $&$Q_{C}$ & $Q_L$ & $Q_1$ &
   $Q_2$ & $Q_3$ & $Q_{4}$ & $Q_{5}$ & $Q_6$ &
   $ SU(2)_{1,..,6}^6$ & $Q_{7}$ & $Q_{8}$\\
   $$ & $$ & $SU(2)$ & $$ & $$ & $$ & $$ & $$ & $$ & $$ & $$ & $$  & $$ & $$ \\
\hline
$F_1$ & $1+b_2+b_4$ & (1,1) & $3\over4$  & $-{1\over2}$ &
   $1\over4$ & $1\over4$ & $1\over4$ &
   ${1\over2}$ & $1\over2$ & $0$ &
   (1,1,2,1,1,1) & $-{1\over2}$ & $-{1\over2}$  \\
$F_2$ & $\beta\pm\gamma$ & (1,1) & $3\over4$  & $-{1\over2}$ &
   $1\over4$ & $1\over4$ & $1\over4$ &
   ${1\over2}$ & $-{1\over2}$ & $0$ &
   (1,1,1,1,1,2) & $1\over2$ & ${1\over2}$  \\
   $F_3$ & $$ & (1,1) & $-{3\over4}$  & ${1\over2}$ &
   $-{1\over4}$ & $-{1\over4}$ & $-{1\over4}$ &
   ${1\over2}$ & $1\over2$ & $0$ &
   (1,1,2,1,1,1) & $1\over2$ & ${1\over2}$  \\
   $F_4$ & $$ & (1,1) & $-{3\over4}$  & ${1\over2}$ &
   $-{1\over4}$ & $-{1\over4}$ & $-{1\over4}$ &
   ${1\over2}$ & $-{1\over2}$ & $0$ &
   (1,1,1,1,1,2) & $-{1\over2}$ & $-{1\over2}$  \\
\hline
 $$ & $1+b_4$ & (1,1) & $3\over4$  & $1\over2$ &
   $-{1\over4}$ & $1\over4$ & $-{1\over4}$ &
   $0$ & ${1\over2}$ & $-{1\over2}$ &
   (1,1,1,2,1,1) & $1\over2$ & $-{1\over2}$  \\
  $$ & $\pm\gamma$ & (1,1) & $-{3\over4}$  & $-{1\over2}$ &
   ${1\over4}$ & $-{1\over4}$ & ${1\over4}$ &
   $0$ & ${1\over2}$ & $-{1\over2}$ &
   (1,1,1,2,1,1) & $-{1\over2}$ & ${1\over2}$  \\
 \hline
 $$ & $b_3+b_4$ & (1,1) & $3\over4$  & $1\over2$ &
   $1\over4$ & $-{1\over4}$ & $-{1\over4}$ &
   $1\over2$ & $0$ & $-{1\over2}$ &
   (1,2,1,1,1,1) & $-{1\over2}$ & ${1\over2}$  \\
 $$ & $\pm\gamma$ & (1,1) & $-{3\over4}$  & $-{1\over2}$ &
   $-{1\over4}$ & ${1\over4}$ & ${1\over4}$ &
   $-{1\over2}$ & $0$ & $-{1\over2}$ &
   (1,2,1,1,1,1) & ${1\over2}$ & $-{1\over2}$  \\
\hline
 $$ & $b_1+b_2+b_3$ & (1,1) & $3\over4$  & $1\over2$ &
   $-{1\over4}$ & $1\over4$ & $-{1\over4}$ &
   $0$ & $-{1\over2}$ & $-{1\over2}$ &
   (1,2,1,1,1,1) & $-{1\over2}$ & $1\over2$  \\
 $$ & $+b_4\pm\gamma$ & (1,1) & $-{3\over4}$  & $-{1\over2}$ &
   ${1\over4}$ & $-{1\over4}$ & ${1\over4}$ &
   $0$ & $-{1\over2}$ & $-{1\over2}$ &
   (1,2,1,1,1,1) & ${1\over2}$ & $-{1\over2}$  \\
\hline
\end{tabular}
\label{matter33}
\end{eqnarray*}

\end{table}

\end{document}